\documentclass[11pt,a4paper]{emulateapj}
\usepackage{epsfig}
\usepackage{amsmath}
\usepackage{natbib}

\begin{document}

\shortauthors{Anjali J Kaithakkal et al.}
\shorttitle{Photospheric Flows and Polar Magnetic Patches}

\title{Photospheric Flow Field Related to the Evolution of the Sun's Polar Magnetic Patches Observed by Hinode SOT}

\author{Anjali John Kaithakkal}
\author{Y. Suematsu}
\author{M. Kubo}
\affil{Department of Astronomical Science, Graduate University for Advanced Studies (SOKENDAI), Mitaka, Tokyo 181-8588, Japan; anjali.johnk@nao.ac.jp.}
\affil{National Astronomical Observatory of Japan (NAOJ), Mitaka, Tokyo 181-8588, Japan.}
\author{Y. Iida}
\affil{Institute of Space and Astronautical Science (ISAS), Japan Aerospace Exploration Agency (JAXA), Sagamihara, Kanagawa 252-5210, Japan.}
\author{D. Shiota}
\affil{Solar-Terrestrial Environment Laboratory, Nagoya University, Nagoya 464-8601, Japan.}
\author{S. Tsuneta} 
\affil{Institute of Space and Astronautical Science (ISAS), Japan Aerospace Exploration Agency (JAXA), Sagamihara, Kanagawa 252-5210, Japan.}


\begin{abstract}

We investigated the role of photospheric plasma motions in the formation and evolution of polar magnetic patches using time-sequence observations with high spatial resolution. The observations were obtained with the spectropolarimeter on board the Hinode satellite. From the statistical analysis using 75 magnetic patches, we found that they are surrounded by strong converging, supergranulation associated flows during their apparent life time and that the converging flow around the patch boundary is better observed in the Doppler velocity profile in the deeper photosphere. Based on our analysis we suggest that the like-polarity magnetic fragments in the polar region are advected and clustered by photospheric converging flows thereby resulting in the formation of polar magnetic patches. Our observations show that, in addition to direct cancellation magnetic patches decay by fragmentation followed by unipolar disappearance or unipolar disappearance without fragmentation. It is possible that the magnetic patches of existing polarity fragment or diffuse away into smaller elements and eventually cancel out with opposite polarity fragments that reach the polar region around solar cycle maximum. This could be one of the possible mechanisms by which the existing polarity decay during the reversal of the polar magnetic field.

\end{abstract}

\keywords {Sun: convective motions -- Sun: magnetic fields -- Sun: photosphere}

\section{Introduction}

The Sun's polar caps are dominated by unipolar magnetic patches which possess magnetic fields in kilogauss range \citep{tsa}. Though polar flux is believed to be originated from surplus magnetic flux of the decayed active regions, it still remains unexplored as to how: a) magnetic flux in the polar region is concentrated in the form of unipolar magnetic patches, and b) these magnetic patches decay and eventually reverse the polarity of the polar field. 

Formation of magnetic structures by transportation and accumulation of magnetic flux driven by converging horizontal flows are observed in the lower heliographic latitudes \citep[e.g.,][]{lin,yi}. Most of the magnetic flux outside sunspots is concentrated and organized into a variety of multi-scale magnetic features (e.g., network and internetwork magnetic structures) by convective flows in the solar surface layers. Horizontal converging flows concentrate vertical magnetic flux predominantly at the convective cell boundaries. The magnetic flux is advected to the cell boundaries until the field strength reaches the equipartition value which corresponds to the balance between magnetic pressure and dynamic pressure of the convective flows. Further intensification of magnetic fields to kG strengths is induced by the mechanism of convective collapse \citep{par,spr}.  

Magnetic fields and photospheric plasma motions are well coupled and hence it is important to obtain detailed information on the role of the flow field in the formation and evolution of the polar magnetic patches. This information might give some insight to understand the mechanism involved in polar field reversal and the dynamical processes that could influence the overlying atmospheric layers. The motivation for this study is the observation, with the Hinode satellite \citep{ks}, of isolated appearance and disappearance of unipolar magnetic patches in the polar region which does not support emerging flux scenario and cancellation with opposite polarity patch respectively. We also wanted to understand whether the time of observation has any role in detecting polar facula inside the patch or not. In this study we investigate the role of photospheric flow fields in the formation and evolution of polar magnetic patches. We used high spatial resolution observations obtained with the Spectropolarimeter \citep[SP;][]{lit} of Solar Optical Telescope \citep[SOT;][]{ichi,tshi,sue,tsb} on board Hinode for this study. Section 2 describes our observation and analysis. The main results obtained are detailed in section 3 and summary and discussion on the results are given in section 4. 


\section{Observation and Data Analysis}

The data sets used in this study are given in Table 1. These data were taken from the north and south polar caps of the Sun with SP. The SP recorded full Stokes spectra of the two Fe I lines at 630.15 nm and 630.25 nm with the fast map mode whose slit-scanning step is 0$\farcs$32 and integration time in each step is 3.2s. The slit was along N - S direction. The spatial sampling along the slit is 0$\farcs$32 and the spectral sampling is 2.15 pm. The FOV of the SP image sequences was 80$\arcsec$ x 164$\arcsec$ and the cadence was 16 minutes. The observations were taken in such a way that the solar limb and the pole is always within the FOV (Figure 1). Consecutive SP image frames are then aligned using spatial cross-correlation of the Stokes $V$ maps, with pixel accuracy to compensate for the the image motion induced by the correlation tracker device on board the SOT. The spatial offsets thus obtained were used to register other relevant parameters. 

\begin{deluxetable*}{ccccccc}
\tabletypesize{\scriptsize}
\tablecolumns{5}
\tablewidth{0pc}
\tablecaption{SP Observations}
\tablehead{
\colhead{Date} &\colhead{Time} &\colhead{Center of FOV of the First Scan} &\colhead{Number of Frames} &\colhead{Number of Patches Selected}\\
\colhead{} &\colhead{(UT)} &\colhead{} &\colhead{} &\colhead{}}\\
\startdata
	
2013 Nov 11	 &10:22-16:43		 &(-15.0$''$, 917.4$''$)		  & 24 &12\\ 
2013 Nov 13	 &09:57-15:51		 &(-14.9$''$, -942.6$''$)		  & 24 & 5\\ 
2013 Dec 08	 &09:00-15:47		 &(-14.8$''$, 917.3$''$)		  & 24 & 10\\ 
2013 Dec 11	 &09:06-15:53		 &(-15.0$''$, -942.5$''$)		  & 24 & 15\\ 
2014 Jan 17	 &09:06-15:53		 &(-14.9$''$, 917.3$''$)	           & 24 & 13\\
2014 Jan 23	 &03:05-09:18		 &(-14.9$''$, -958.0$''$)		  & 23 & 13\\
2014 Mar 08	 &11:06-17:43		 &(-14.9$''$, -957.8$''$)		  & 24 & 7\\

\enddata
\end{deluxetable*}

\begin{figure}[htbp]
  \begin{center}
      \includegraphics[width=6cm,height=10cm]{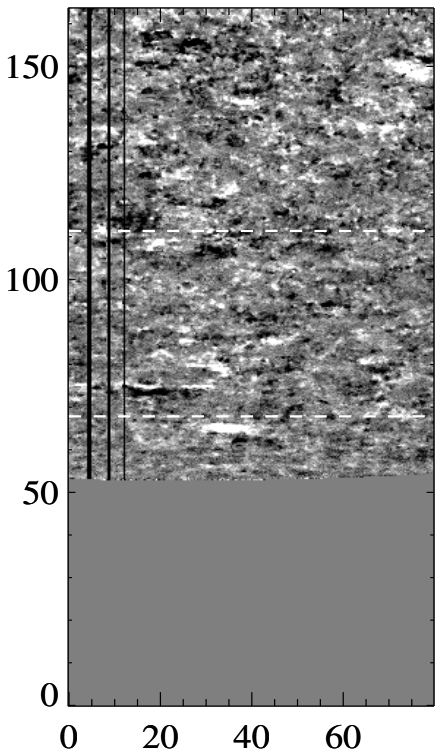}
    \end{center}
   \caption{SP Stokes $V$ map (2014 March 08). The $x$ and $y$ axes are in arcsec. Regions between the dashed lines correspond to our region of interest (latitude band of 70$^{\circ}$ - 80$^{\circ}$). }
    \end{figure}
    
To determine the solar pole position, a filtergram (NFI/SOT) image taken at time close to the time of scanning of the center of FOV of the first frame of the SP observations was used. The FG data (shuttered Stokes $I$ and $V$) were taken at 140 m{\AA} red-ward of the Na {\sc i} D$_1$ 5896{\AA} line center with a FOV of 327$\farcs$7 x 163$\farcs$8. The pixel size was 0$\farcs$16 and the the exposure time was 0.205 s. The solar limb position was estimated using the FG Stokes $I$ image and was then fitted with a circle to calculate the pole positions. We used large FOV FG image to obtain solar limb as it is required to minimize error while fitting the limb with a circle to derive the pole position. FG Stokes $V$ image and SP Stokes $V$ (first scan of the observation) map were then aligned using spatial cross-correlation to get the information on the limb and pole positions in the SP map scale. This information is used to derive $\mu$ (cosine of the heliocentric angle) which is used to obtain the normalized intensity (${I_c}/{\langle I_c \rangle}$). 
     
The magnetic field parameters were derived by using a least-squares fit to the Stokes profiles using the MILOS code \citep[\textbf{Mil}ne-Eddington Inversion of P\textbf{o}larized \textbf{S}pectra;][]{do}. MILOS  assumes  a two-component atmosphere model (magnetic and nonmagnetic) in a pixel. The inversion code provides: three components describing the vector magnetic field - the field strength $B$, the inclination between the line-of-sight (LOS) and the field vector, and the azimuth of the field vector in a plane perpendicular to the LOS, the LOS velocity, two parameters describing the linear source function, the ratio of line to continuum absorption coefficients, the Doppler width, the damping parameter, and the stray-light factor $\alpha$. The stray-light factor, $\alpha$ quantifies contribution to the measured intensity from both a non-magnetic area of the pixel and stray light contamination arising from instrument optics. If stray light contamination is negligible, magnetic fill fraction - fraction of a pixel occupied by magnetic field - is calculated as $f = 1 - \alpha$ \citep{doa}. 

The inversion is performed only for pixels whose linear or circular polarization signal amplitudes exceed a given threshold above the noise level which depends on the exposure time. The noise level $\sigma'$ is determined in the continuum wavelength range of the Stokes V profiles and is given by $\sqrt{\sum_i (V_i - \overline{V} )^2/n)}$, where $V_i$ is the intensity of the Stokes $V$ profile at continuum wavelength pixel $i$, $\overline{V}$  is the average Stokes $V$ signal for the same wavelength range, and $n$ is the number of wavelength data points. The noise level $\sigma'$ is 1.3 x 10$^{-3}I_c$, where $I_c$ is the continuum intensity. The pixels with Stokes $Q, U$ or $V$ peak larger than 5$\sigma'$ alone are fitted using the code.

The azimuth value provided by the inversion is ambiguous by 180$^{\circ}$. This ambiguity in the transverse magnetic fields was resolved by employing the method of \citet{ito}. The vector magnetic field for each pixel  will have two solutions for the local zenith angle as a result of the ambiguity. We assume that the magnetic field vector is either vertical or horizontal to the local solar surface. If the zenith angle is between 0$^{\circ}$ and 40$^{\circ}$ or between 140$^{\circ}$ and 180$^{\circ}$, the field vector is taken to be vertical. Magnetic field vectors associated with polar patches are nearly vertical to the local solar surface \citep[e.g.,][]{tsa}. We define vertical magnetic flux as $\sum_jB_j f_j \cos i_j A_j$, where $B_j$, $f_j$, $i_j$, and $A_j$ are the intrinsic field strength, magnetic filling factor, local zenith angle, and pixel area, respectively, for the $j$th CCD pixel. The pixel area was corrected for projection effect.  
     
\subsection{Identification and Tracking of Magnetic Patches}
The code developed by \citet{iida}, which was used to identify and track network magnetic patches, in quiet Sun near disk center (for details see Iida 2012, Section 2.2.2) is employed here to select and track the polar magnetic patches. The modifications from Iida$'$s code and conditions used in this study are detailed as follows. Magnetic patches must:\\
(1) be within the heliocentric latitude band of 70$^{\circ}$ - 80$^{\circ}$\\
(2) have minimum size of 5 contiguous  pixels\\ 
(3) have per pixel flux greater than 2 x 10$^{16}$ Mx (1 sigma threshold value obtained from the magnetic flux map)\\
We identify patches from the magnetic flux map with a clumping method. The clumping method chooses and groups all connected pixels which satisfy the above criteria into a single magnetic patch.

Here we consider the lateral shift due to rotation ($\propto$ sin$\phi$, where $\phi$ is the colaitude) is small within the latitude range of 70$^{\circ}$ - 80$^{\circ}$ \citep[see,][]{ben} and granular advective velocity of 1km/s \citep[e.g.,][]{berg}. Thus the magnetic patches are assumed to undergo a maximum displacement of about 4 pixel size within an interval of 16 minutes (1 km/s x 960 s = 960 km $\sim$ 4 pixels). The magnetic patches which spatially overlap in consecutive SP frames are marked as identical. Those samples which were born and disappeared during the period of observation, with minimum life time of 3 frames are chosen. We eliminated patches that are located close to the edge of the FOV. Finally, 75 magnetic patches in total satisfied the above criteria. 

We identified facular pixels inside each patch in the normalized continuum intensity maps.  A smoothed map representing the center-to-limb variation (CLV) of the continuum intensity, $\langle I_c \rangle$, is obtained through a two step process: A least-squares surface fit using a 5th-order polynomial in $\mu$ \citep[following][]{pi} is performed on the $I_c$ map. The fitted map is then subtracted from the original, and the standard deviation $\sigma_{0}$ of the difference is calculated. We then removed the bright and dark features from the original image using a $\pm3\sigma_{0}$ cutoff, and a fit with same functional form is performed to obtain a CLV function unaffected by the presence of faculae and dark features. The normalized intensity is defined as ${I_c}/{\langle I_c \rangle}$, where $I_c$ and $\langle I_c \rangle$ are the continuum intensity and the intensity averaged over the same $\mu$-value, respectively. 

Within each magnetic patch, pixels having intensity greater than or equal to a given threshold are classified as belonging to polar faculae. The threshold to identify facular pixels varies with $\mu$. The standard deviation $\sigma$ of the normalized intensity is derived at each $\mu$ with a bin size of 0.01, and the $\sigma$'s are fitted with a 3rd-order polynomial in $\mu$. The threshold to detect faculae is set to 3 $\sigma$.

\subsection{Bisector Analysis}

Visual inspection of the magnetograms obtained with the SP observation show unipolar appearance and disappearance of the polar magnetic patches. We investigated whether the photospheric flow field around the patches has any role in the appearance and disappearance of the polar magnetic patches. The  inversion does not yield height dependent LOS Doppler velocity. The variation of the flows with height is examined by the bisector analysis of the Fe {\sc i} 630.15 nm line profile. This spectral line is less sensitive to the magnetic field (g = 1.67) in comparison with the Fe {\sc i} 630.25 nm (g = 2.5) line. Though the magnetic sensitive line is used, the effect of magnetic field on the velocity measurements is assumed to be negligible in the nearly field-free plasma surrounding the magnetic patch. We obtained bisector positions of the line profile at four intensity levels between line core and wing (see Figure 2). The formation height decreases with increase in intensity along the line profile. Thus, the bisector level 4, shown in Figure 2, forms deeper in the solar photosphere than the bisector level 1. The Doppler velocity is calculated as v = ($\Delta\lambda/\lambda_0$) c, where $\lambda_0$ is 630.15 nm and $c$ is the velocity of light. 

\begin{figure}[htbp]
  \begin{center}
      \includegraphics[width=9cm,height=6cm]{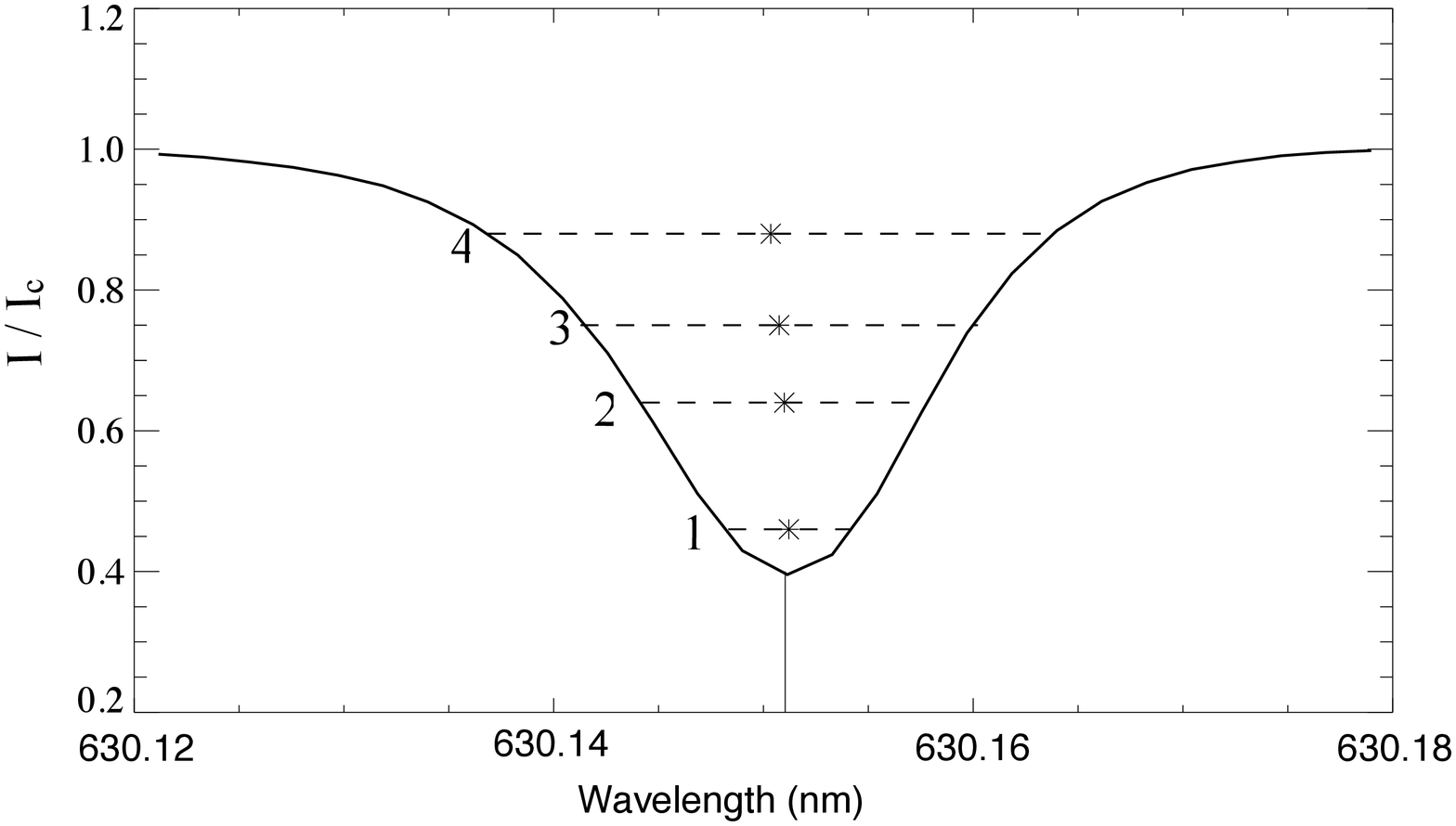}
       \caption{Normalized Stokes $I$ profile of the Fe {\sc i} 6301.5 {\AA} absorption line. The solid vertical line represents the line core position and the asterisks represents the bisector points.}
\end{center}
    \end{figure} 

The Doppler shift for each bisector level ($\Delta\lambda$) is determined with respect to the reference wavelength position at that level. As we do not have an absolute reference wavelength position, the reference wavelength is determined as follows. For each of the selected sample patch we define a vertical (north-south) slot of height about 96$''$ in the slit direction, excluding pixels close to the limb, and width same as that of the patch (defined as the difference between maximum and minimum locations of the patch across the slit direction). The spectral line profiles in this vertical slot were then averaged to obtain a mean spectral line profile. The reference wavelength position for each of the bisector levels was calculated at the respective intensity positions from the mean spectral profile. The reference wavelength position at each bisector level was found to vary by about $\pm$0.2 pm ($\sim$ 0.1 km/s) between image sequences in which a given patch is present .  

Since we are interested in a relative velocity in the region around the magnetic patches, we defined a reference wavelength which gives an average velocity in the region of our interest. In this study, we defined a sub vertical slot of width same as that of the patch and height $\pm$8$''$ from the top and bottom boundary respectively of each patch (for e.g., see Figure 3). The velocity averaged over this sub slot is defined as zero (reference) velocity in our study.

 \begin{figure}[htbp]
  \begin{center}
      \includegraphics[width=6cm,height=7cm]{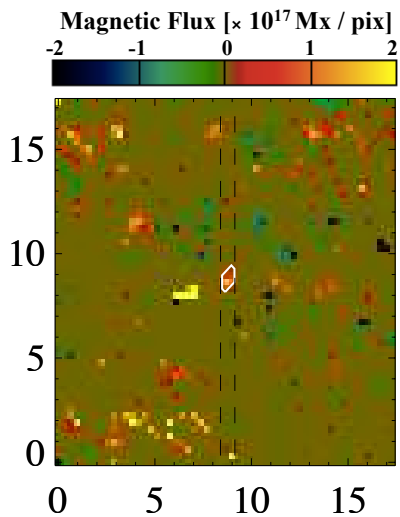}
       \caption{Magnetic flux map of a sample patch (2014 January 23) at time $t_0$ - time corresponding to the first appearance of the magnetic patch. The solid line encircles the boundary of the patch. The dashed lines mark the edges of the sub vertical slot mentioned in the text. The $x$ and $y$ coordinates are in arcsec. This is one of the samples which clearly show a gradient in velocity outside of the magnetic patch in the slit direction.}
\end{center}
    \end{figure}


\section{Results}

\subsection{Lifetime and Magnetic Flux Distribution of the Samples}

Here, we outline the general properties like lifetime and magnetic flux distribution of the 75 magnetic patches chosen as described in Section 2.1. The distribution of apparent life-time of the sample magnetic patches is shown in Figure 4. Most of the samples have a life time of 32 min (3 frames) and the average life time is about 1 h. Figure 5 shows the distribution of time-averaged magnetic flux of the patches. The average magnetic flux is $\sim$ 10$^{18}$ Mx. This value is close to the lower limit of the large flux concentration mentioned in \citet{shi}. Majority of the patches have positive polarity since patches with positive polarity are dominant in both the north and the south polar caps during our observation period. The patches with negative polarity (22 patches) come from both north and south polar region. There are many magnetic patches with larger flux which were present during the entire observation period (6 h) and are not considered in this study. 

\begin{figure}[htbp]
  \begin{center}
      \includegraphics[width=8cm,height=9cm]{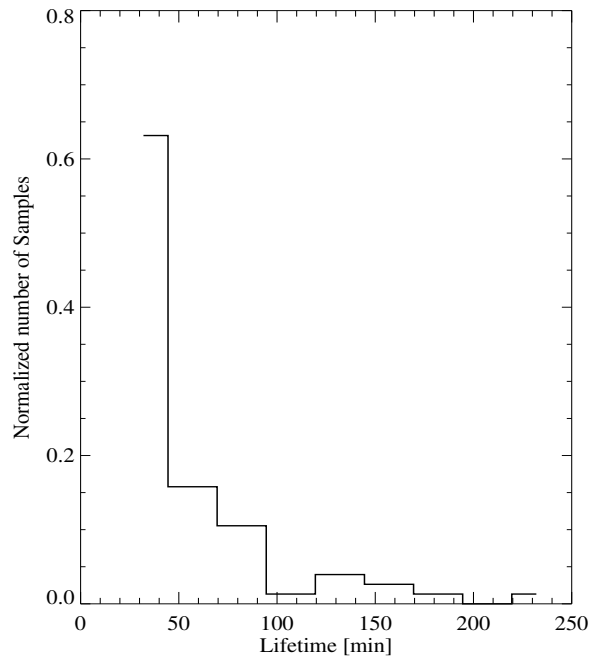}
    \end{center}
   \caption{Distribution of lifetime for the 75 sample patches selected for this study.}
    \end{figure}

\begin{figure}[htbp]
  \begin{center}
      \includegraphics[width=9cm,height=7cm]{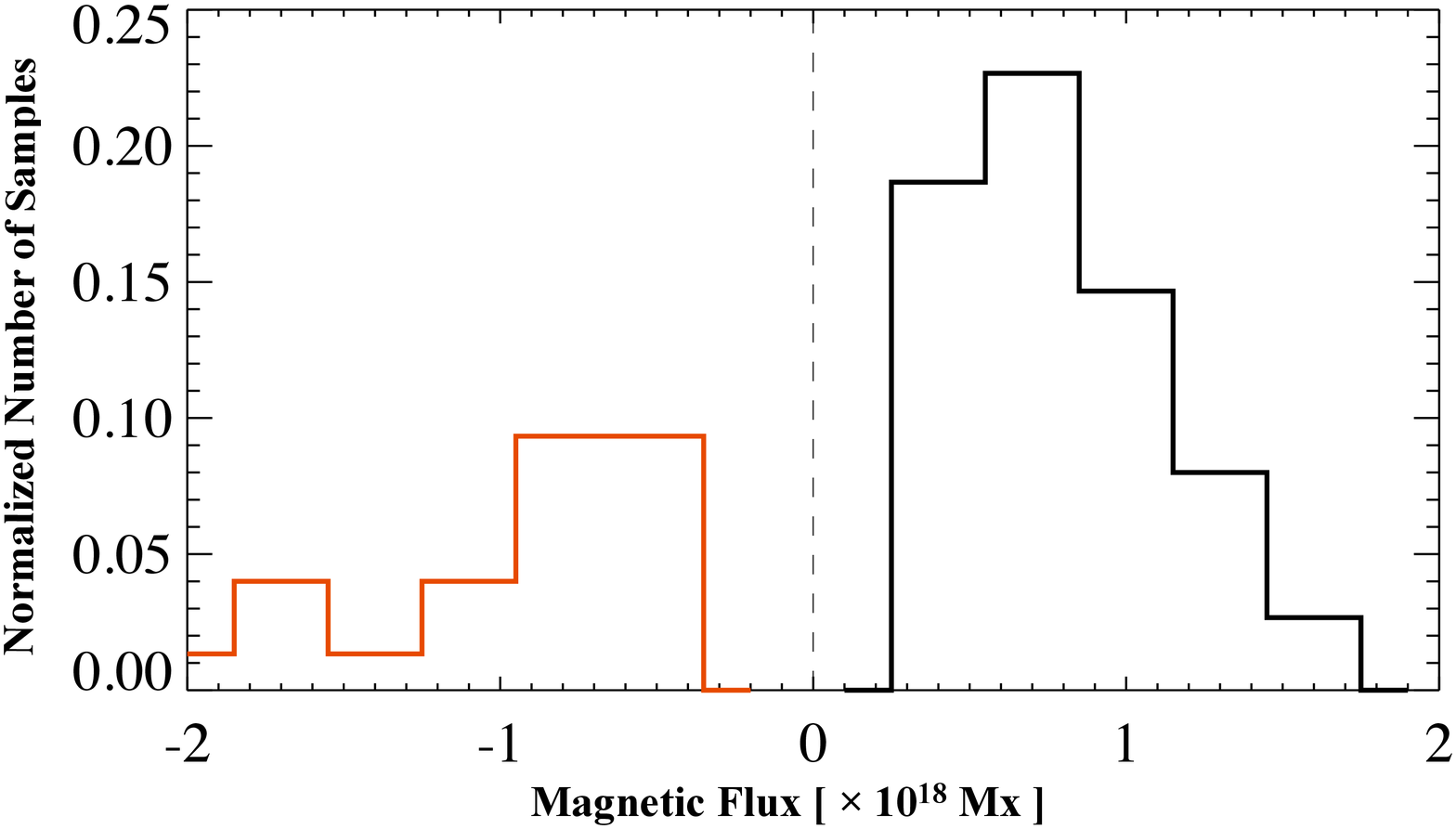}
    \end{center}
   \caption{ Histogram of time-averaged magnetic flux of the 75 samples chosen for this study. Distribution of negative polarity patches (minority polarity) is plotted in red color and  positive polarity patches (majority polarity) in black color.}
    \end{figure}
    
\subsection{Flow Field at the Time of Appearance and Disappearance of Magnetic Patches}
 
 \subsubsection{Appearance of Magnetic Patches}                                                                                                                                                                                                                     
 
Here we discuss the photospheric flows in and around the 75 magnetic patches during their appearance. The time at which a magnetic patch is detected for the first time, is termed as $t_0$. The velocity profile in and around the patch along the slit direction is obtained as follows. To minimize the effect of noise, the velocity at each position (within the sub vertical slot) along the slit direction is obtained by averaging the Doppler velocity over the width of the sub vertical slot across the slit direction. In general, magnetic patches have a 'ragged' shape and hence have non-uniform width across the patch. So if the width of the patch is smaller than the width of the slot at a given location within the patch, the average is calculated only over  those positions within the patch. This separate treatment for the magnetic patch is performed to understand the nature of the flow velocity in the presence of the magnetic field. The reference velocity is subtracted from the Doppler value obtained at each position along the slit. A sample velocity profile at the bisector level 4 at time $t_0$ is shown in Figure 6. The zero position on the $x$- axis is the location within the patch at which the average intensity becomes maximum. The velocity profile shows dominance of blue shift on the limb-ward side and red shift on the disk center-ward side within a distance of $\pm$ 2$''$ respectively from the patch boundary. Blue- and red-shifted flows on the limb- and disk center-ward directions respectively of the patch represent the existence of converging (incoming) flow field. For each patch, we retraced the patch location at time $t_0$ onto the frame at $t_0$-16 min (Figure 7). Doppler velocity in the sub vertical slot at $t_0$-16 min is determined using the same method as explained before to examine whether the flow field exhibit any trend prior to the patch appearance around the retraced location. The precursor was not always observed in the magnetograms at $t_0$-16 min.

\begin{figure}[htbp]
  \begin{center}
      \includegraphics[width=9cm,height=6cm]{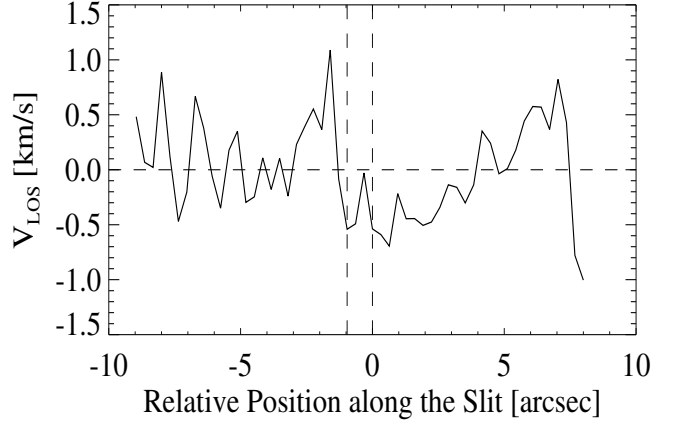}
   \caption{The Doppler velocity profile for the magnetic patch shown in Figure 3. The position where the average of the normalized intensity becomes maximum within the magnetic patch is defined as 0 in the $x$ axis and the limb is toward right. The vertical dashed lines represent edges of magnetic patch on its limb-ward and disk center-ward side. Positive velocities correspond to flows away from the observer (redshift). The $\mu$ value of the magnetic centroid of the patch is 0.31.}
    \end{center}
    \end{figure}
    
 \begin{figure}[htbp]
  \begin{center}
      \includegraphics[width=6cm,height=6.5cm]{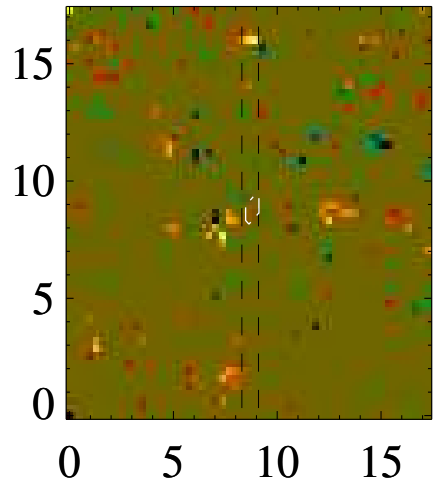}
    \end{center}
   \caption{The location of the patch in Figure 3 is retraced to the frame at $t_0$-16 min.The dashed contour represents the location of the patch at time $t_0$.}
    \end{figure}

The above procedure is carried out for all 75 samples to get an average flow field around the retraced patch location at $t_0$-16 min at the four bisector levels. Figure 8 shows weak converging flow around the retraced location of the patch at $t_0$-16 min. Average Doppler profiles for regions within and around patches at time $t_0$ is shown in Figure 9. The plots on the left display that the patch is surrounded by systematic converging flow at all the four bisector levels. The velocity profiles within the patch shows that converging flow continues more or less within the patch. The slight difference in the flow continuity could be due to the difference in two regions: one magnetic and the other nearly non-magnetic. The redshift becomes weaker in the higher layers but the blueshift does not change with height. Considering the patch and its surrounding together, it appears that the horizontal flow is converging to the zero position. 

  \begin{figure}[htbp]
    \begin{center}
 \includegraphics[width=7.5cm,height=16cm]{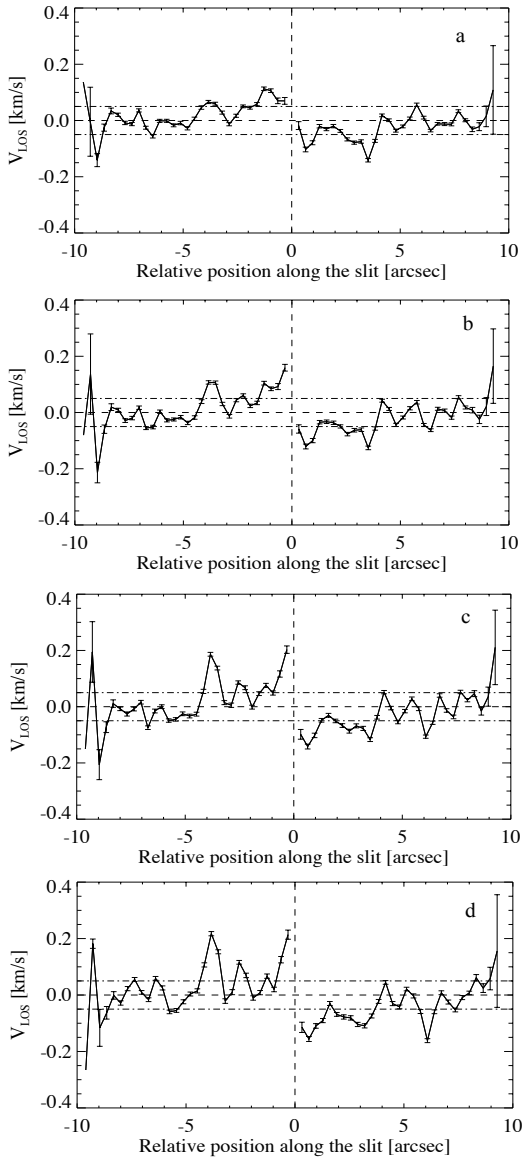} 
    \caption{The average Doppler velocity profiles at bisector levels: (a) 1, (b) 2, (c) 3, and (d) 4 at time $t_0$-16 min, from top to bottom. The solid vertical lines represent measured standard errors. The horizontal dash-dotted lines represent $\pm$1 standard deviation value obtained from the non-magnetic region(see text for details). Limb is towards right. Positive velocities correspond to flows receding from the observer. To obtain this, all velocity profiles from the south polar region are rotated to have the direction of the limb point northward.}
\end{center}
\end{figure}  
        
\begin{figure*}[htbp]
\centering
   \includegraphics[width=12cm,height=14cm]{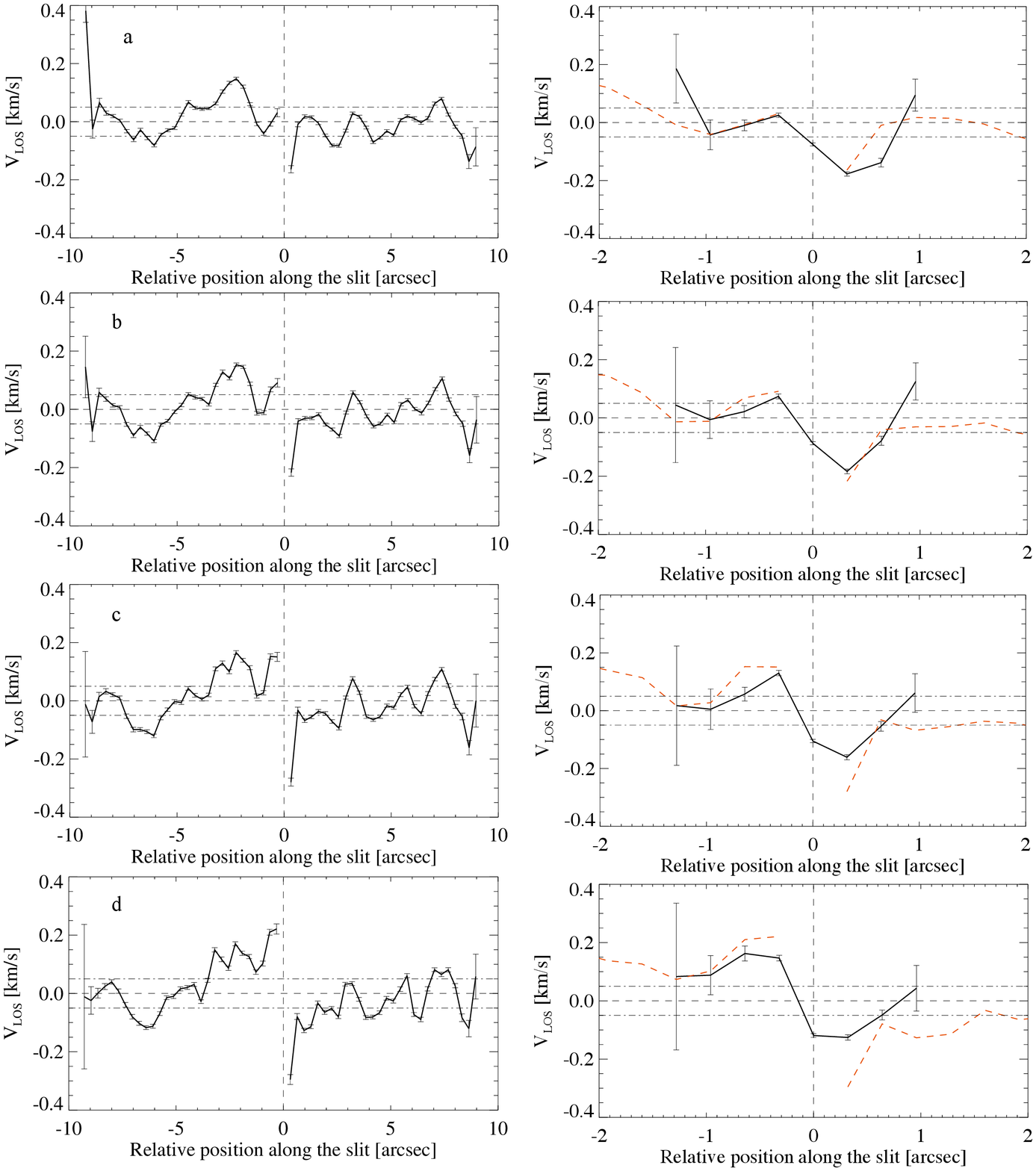}  
   \begin{center}
    \caption{The average Doppler velocity profiles at bisector positions: (a) 1, (b) 2, (c) 3, and (d) 4, respectively from top to bottom at time $t_0$. The plots on the left panel represent LOS velocity field outside the patch and those on the right represent the velocity field within the patch.}
\end{center}
\end{figure*} 

The Doppler velocity values on either side of the zero point come from the region outside the patch. The redshift dominate within a distance of 3$''$ ($\sim$ 10$''$ after foreshortening correction, at $\mu$=0.3 ) outside the patch in the disk center direction. Although the extent of the blueshift on the limb side of the patch is not quite clear, it must also be of the same order as the redshift on the disc center-ward side of the patch. 

The horizontal dash-dotted lines in Figures 8 and 9 represent the standard deviation value obtained from the average velocity profile of non-magnetic region. In order to derive this, the spatial mask corresponding to the location of the patch at $t_0$ was obtained and is shifted randomly across the slit direction such that the region within the mask has zero flux. Then, the Doppler velocity within the sub vertical slot was determined exactly in the same manner as was done with the magnetic patches. This procedure is repeated for all samples and their average Doppler velocity profile as well as its standard deviation which is $\pm$ 0.05 km/s was obtained. Figure 10 show the average profile of the non-magnetic region. The average Doppler velocity profile does not exhibit a systematic flow pattern; it rather shows random variations. 

  \begin{figure}[htbp]
  \begin{center}
      \includegraphics[width=7.5cm,height=8cm]{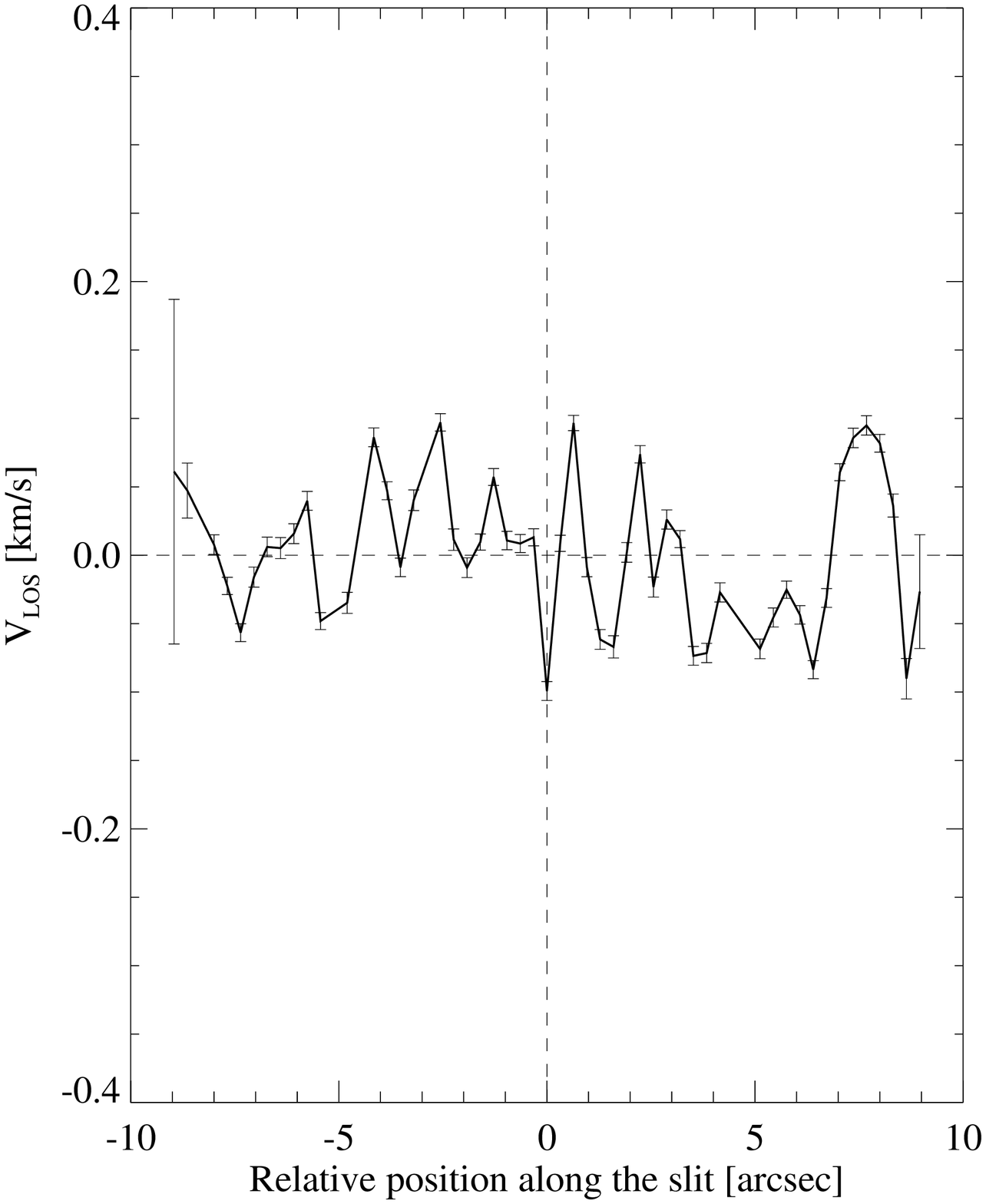}
   \caption{The average Doppler velocity profile of the non-magnetic region. Standard error values are represented by the solid vertical lines.}
     \end{center}
    \end{figure}    

We also calculated average Doppler velocity, at time $t_0$ -16 min and $t_0$, over a distance of 3$''$ from the patch boundary on both the limb- and the disk center-side for each sample. Figure 11 shows distribution of the average velocity on the limb- and the disk center-side of the patch separately. The velocity distribution outside the retraced location of the patch at $t_0$-16 min in the limb ward direction show the existence of an incoming flow toward the patch location, whereas the distribution in the disk ward direction does not exhibit a clear incoming flow. However the histograms show a velocity gradient which indicate the presence of converging flow. The histograms at time $t_0$ support the existence of converging flow around the patch. Although the dominance of redshift on the disk center side of the patch is not that evident from the histogram at time $t_0$, the percentage of samples (specified in the plot) with redshift indicate its prevalence on the disk center-ward direction of the patch.

 \begin{figure}[htbp]
   \begin{center}
   \includegraphics[width=9cm,height=10.5cm]{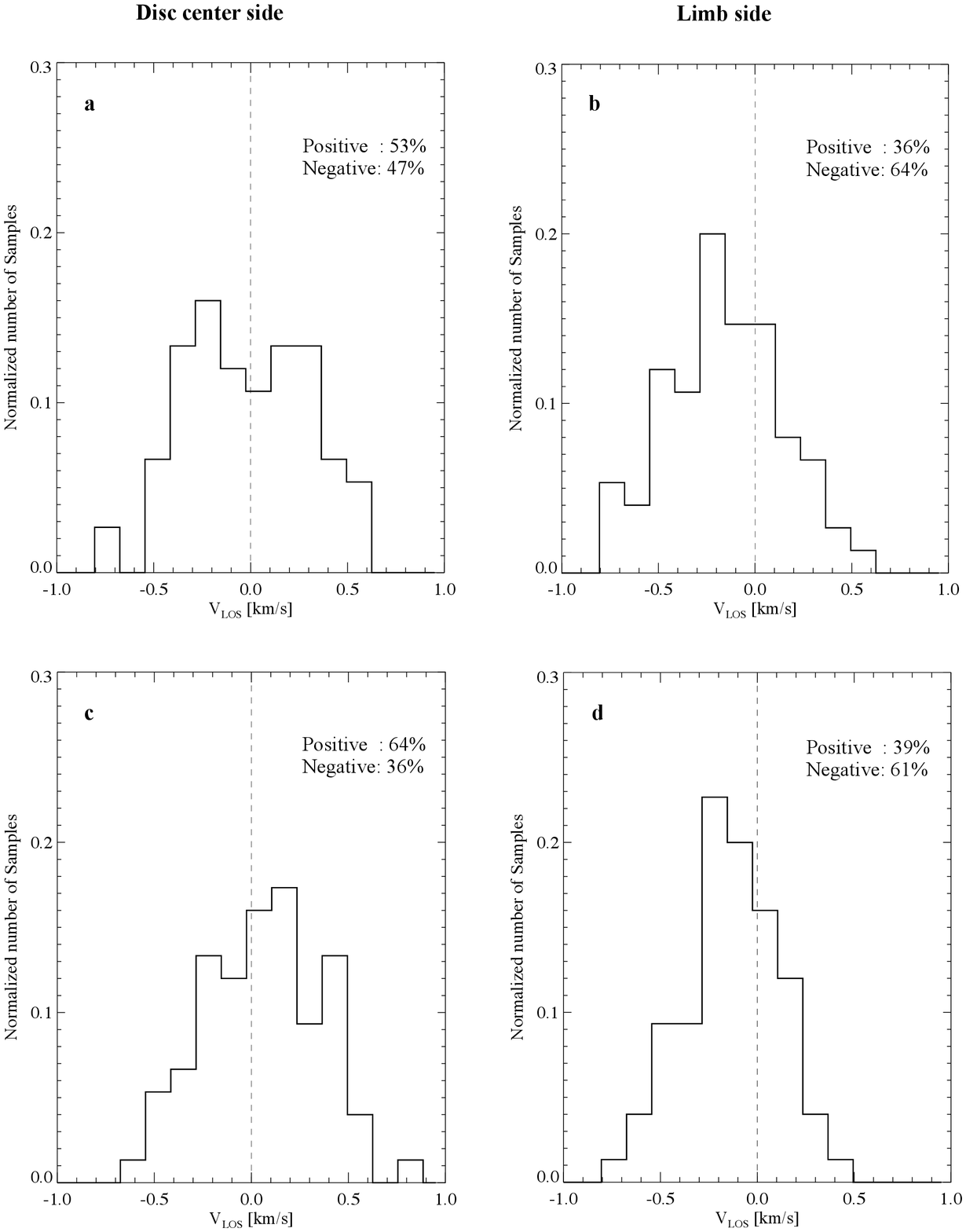}  
    \caption{Panels a and b represent distribution of average Doppler velocity over 3$\arcsec$ from patch boundary at time $t_0$-16 min on the disk center- and limb-side respectively of the patch for the 75 samples. Panels c and d are same as a and b respectively, but for time$t_0$.}
\end{center}
\end{figure}

Inspection of the velocity profiles of the individual samples has shown that not all of them exhibit the converging flow pattern around the patch as shown by the average profile. If the flow field is formed by a few discrete convective cells, the resultant velocity need not always be in a direction favorable to the LOS. This could be one of the reasons for the non-existence of the flow pattern in those samples. 

We define the strength of the converging flow field as the difference in peak velocity at the disk center-  and the limb-ward side of the patch ($V_{\rm r} - V_{\rm b}$) within the 3$''$ zone around the patch. The variation of strength of the converging flow field with bisector levels is shown in Figure 12. The strength of the LOS velocity outside of the patch increases as depth increases. 

   \begin{figure}[htbp]
  \begin{center}
      \includegraphics[width=8cm,height=5.5cm]{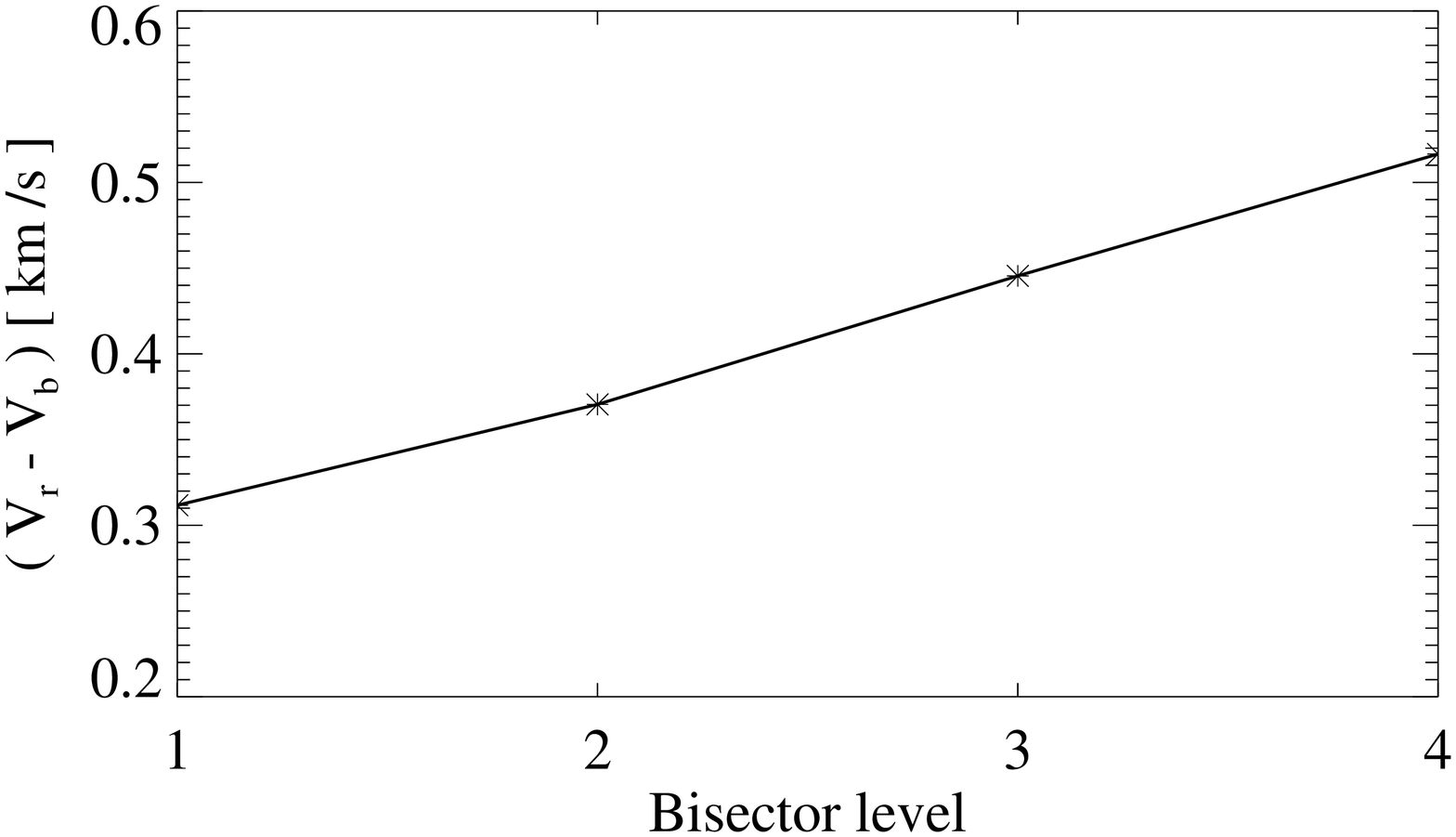}
   \caption{The strength of the converging flow as a function of the bisector level.}
     \end{center}
    \end{figure}    

\subsubsection{Apparent  Death of Magnetic Patches}

The time corresponding to the last frame in which the magnetic patch was visible is defined as $t_{\rm f}$. Figure 13 shows two examples of magnetic patches at time $t_{\rm f}$ and $t_{\rm f}$+16 min. The first sample clearly shows the death of the patch via fragmentation, and the second sample seems to be a case of unipolar disappearance (death in isolation - absence of like- and opposite polarity magnetic features). To investigate whether any trend in flow pattern exists during the apparent death of the magnetic patches, the same procedure described in section 3.2.2 was performed for the 75 samples which were born and disappeared during the period of observation. The average velocity profiles at $t_{\rm f}$ for both the patch and the region surrounding it are shown in Figure 14 The profiles on the left side of the figure show that converging flow exist outside the patches. We found that the converging flow is not continued within the patch and that the flow velocity is redshifted. 

\begin{figure}[htbp]
  \begin{center}
    \includegraphics[width=9cm,height=11cm]{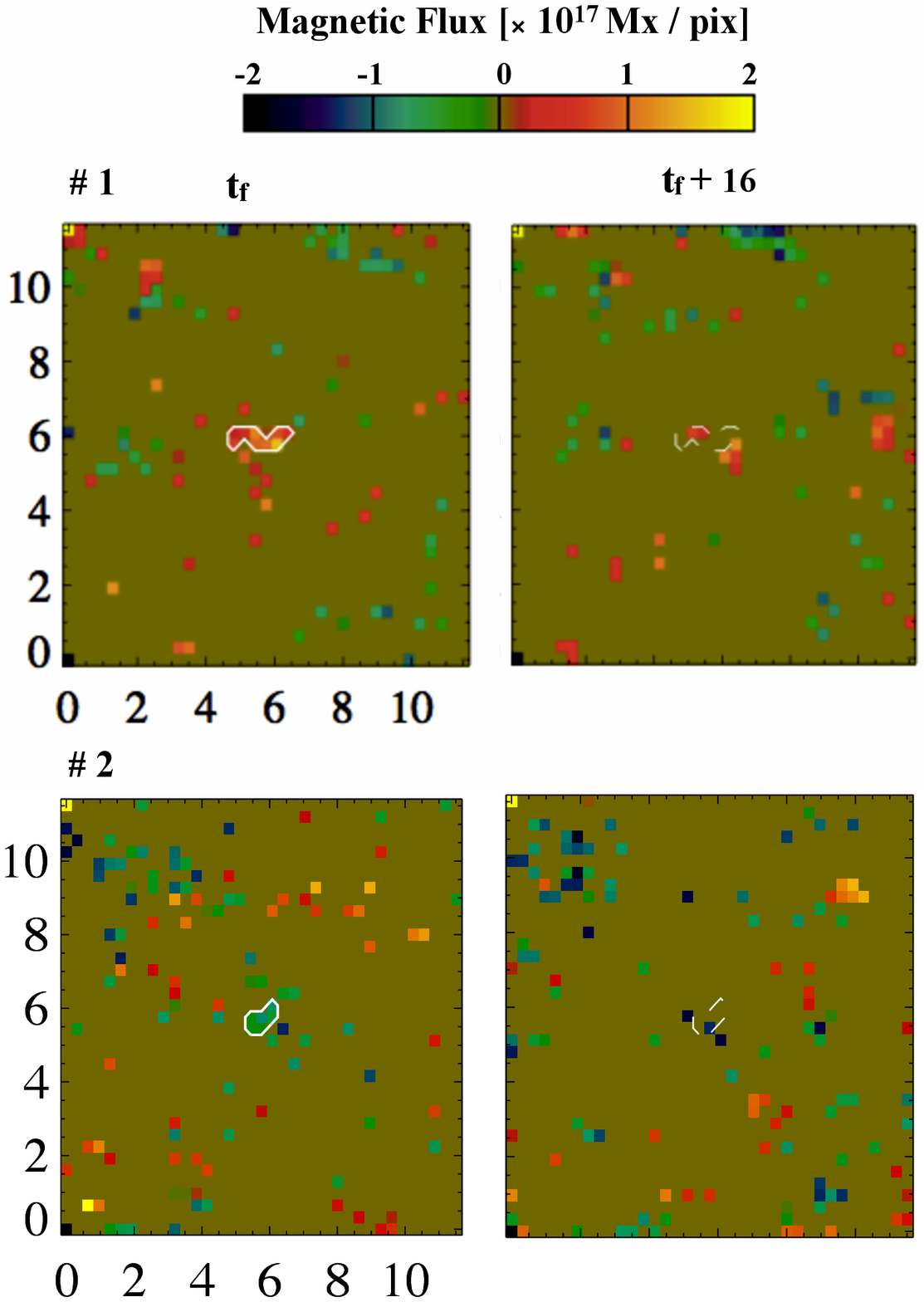}
   \caption{Two examples of disappearance of magnetic patches. The upper panels show magnetic flux maps for sample \#1 (2013 Dec 11) at $t_f$ (left panel) and at $t_f$ + 16 min (right panel). The bottom panels are same but for sample \#2 (2014 January 17).The solid white contour encloses the patch. The dashed contour in right panels represents the spatial location of the patch in the frame at time $t_f$. The $x$ and $y$ coordinates are in arcsec. These two samples show two modes of death of the patches; fragmentation (sample \#1) and unipolar disappearance (sample \#2 ). We didn't classify all the samples based on their mode of death as it is difficult to do so due to lower cadence of the image sequence.}
     \end{center}
    \end{figure}
   
    \begin{figure*}[htbp]
    \centering
    \includegraphics[width=14cm,height=14cm]{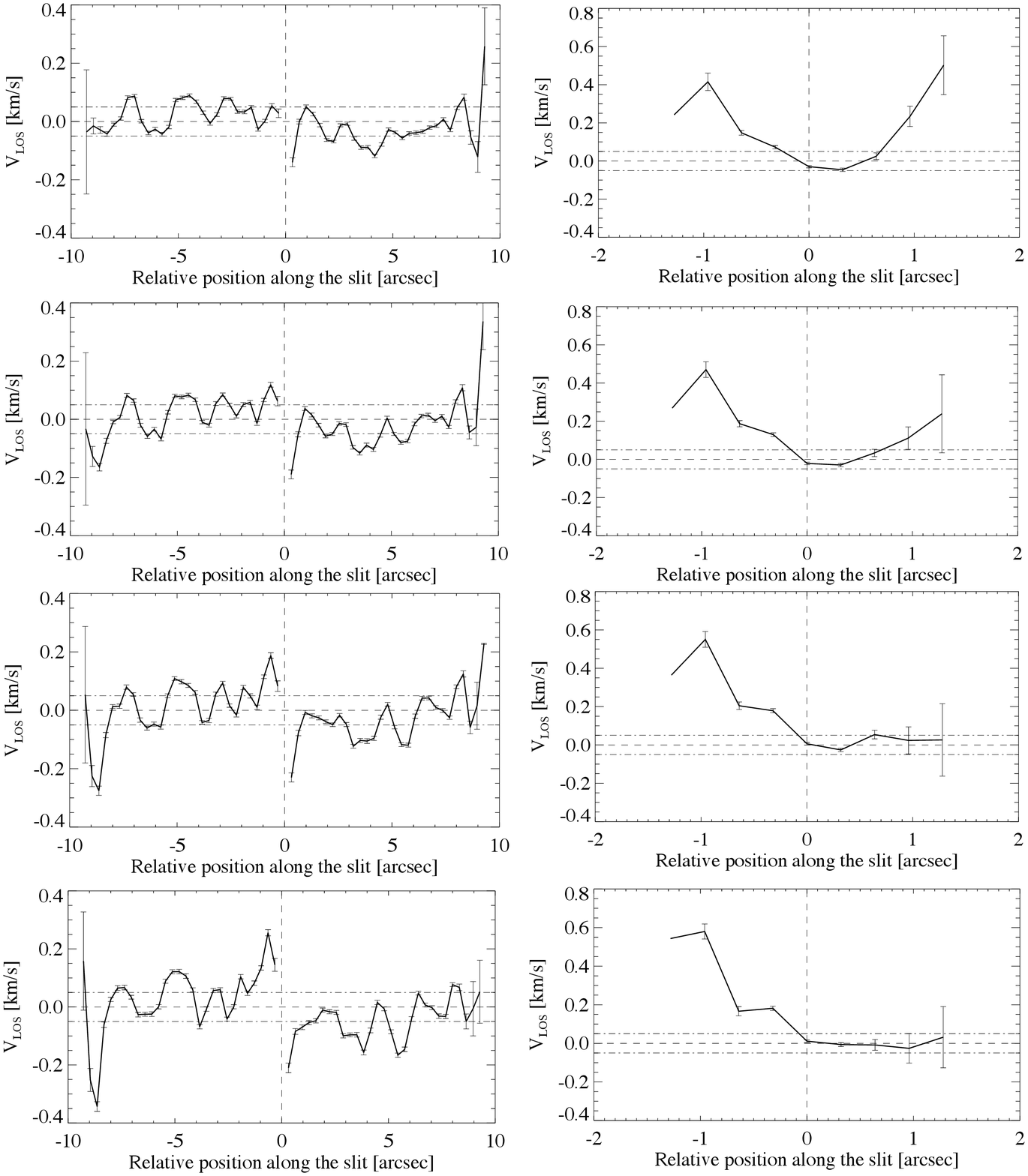}
    \begin{center}
     \caption{Same as Figure 9 but for frames at time $t_f$.}
       \end{center}
    \end{figure*}

The average Doppler velocity, at time $t_f$, over a distance of 3$''$ from the patch boundary on both the limb- and the disk center-side for each sample was determined. Figure 15 shows distribution of the average velocity on the limb- and the disk center-side of the patches separately. The histograms display the existence of incoming flow outside the patch. When compared to velocity distributions at time $t_0$ (Figure 11) the peak of the histograms at time $t_{\rm f}$ is closer to zero. But this small shift is insignificant compared to the standard deviation of the average velocity of the individual samples ($\sim$ 0.3 km/s). Also we found that the difference in the value of $V_{\rm r} - V_{\rm b}$ at bisector level 4 between time $t_0$ and $t_{\rm f}$ is also small. This indicates that converging flow around the patch at time $t_{\rm f}$ is not weak and might of the same order as that at time $t_0$.

 \begin{figure*}[htbp]
   \centering
   \begin{center}
   \includegraphics[width=10cm,height=7cm]{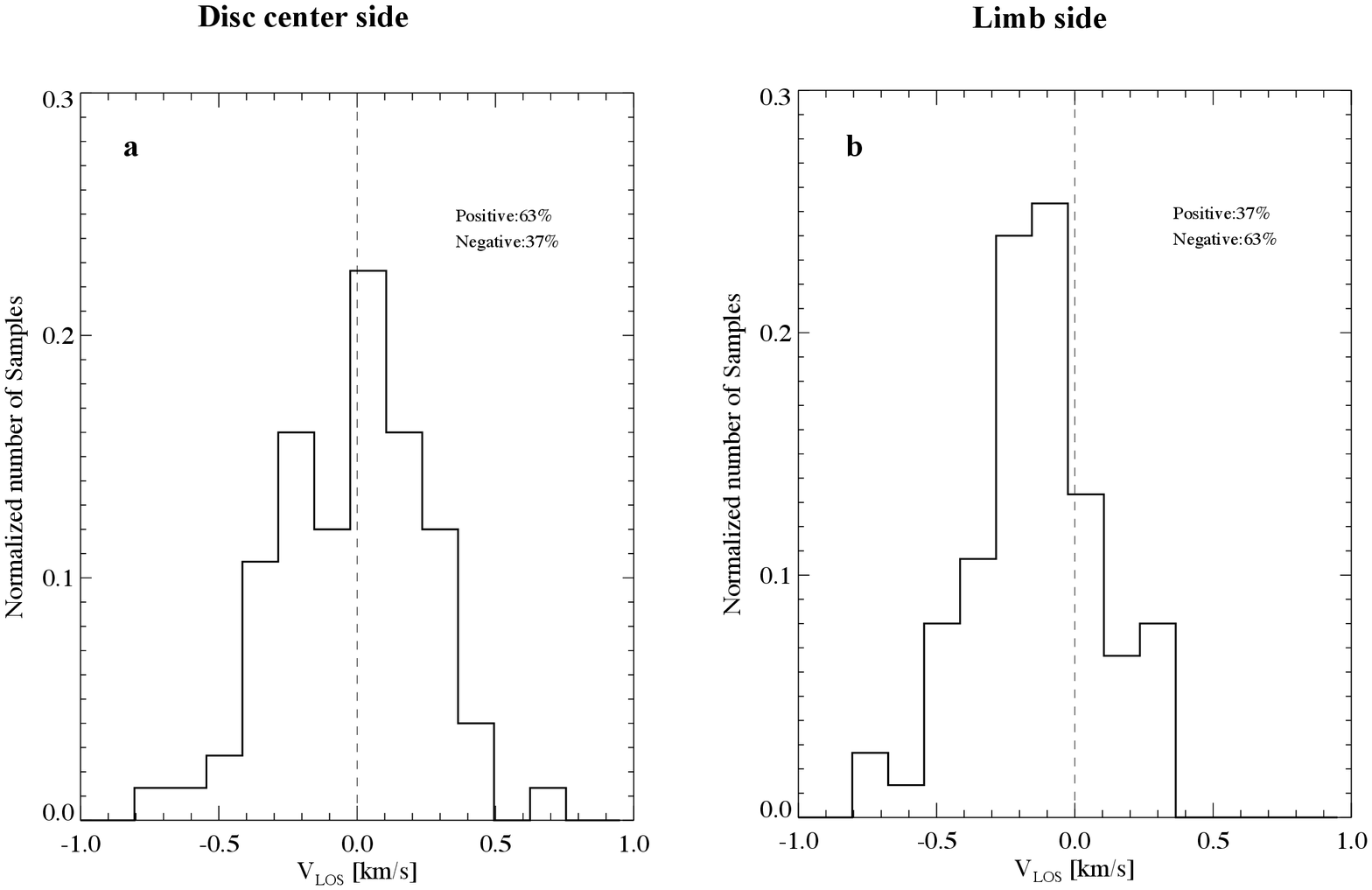}  
    \caption{Panels a and b represent distribution of average Doppler velocity at time $t_f$ on the disk center- and limb-side respectively of the patch.}
\end{center}
\end{figure*}

We also obtained average Doppler velocity profiles at $t_{\rm f}$ + 16 min, at bisector level 4, for the region surrounding the patch along the slit direction which are shown in Figure 16. The profile shows a weak converging flow within a radial distance of 3$''$ outside patches. To verify whether a systematic converging flow exist or not, we obtained distribution of average velocity on the limb- and the disk center-side of the patches separately over a distance of 3$''$ from the patch boundary. The histograms are shown in Figure 17 and the percentage of samples in the distribution is specified in the plot. The distributions show that  there is a relative velocity difference between the limb- and disk center ward sides outside of patches which suggests the existence of converging flow at time $t_{\rm f}$ + 16 min.

 \begin{figure}[htbp]
    \begin{center}
   \includegraphics[width=8cm,height=5.5cm]{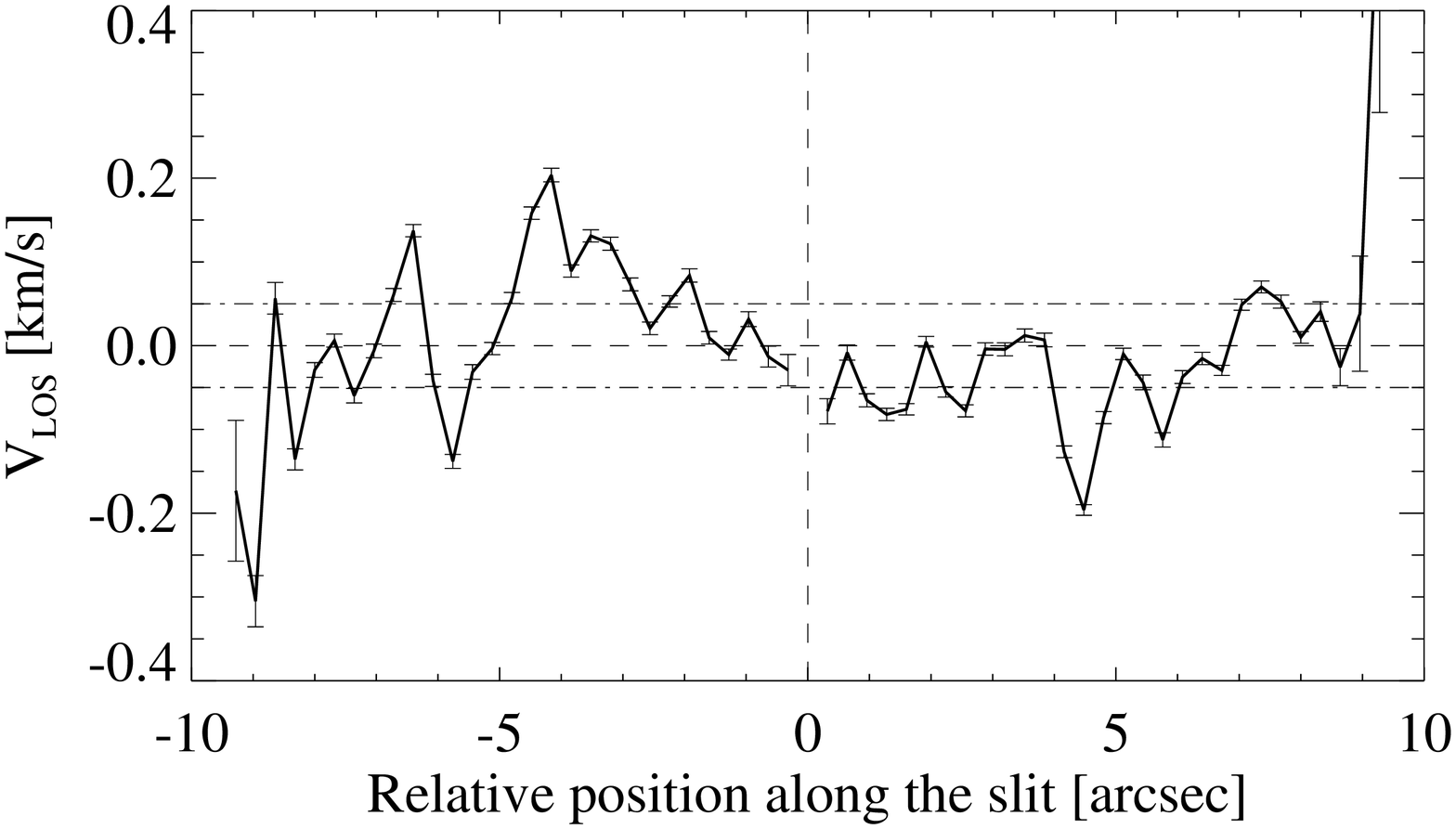}
     \caption{The average Doppler velocity profile at bisector level 4 for the frame at time $t_f$ + 16 min.}
     \end{center} 
    \end{figure}
 
 \begin{figure}[htbp]
   \begin{center}
   \includegraphics[width=9cm,height=6cm]{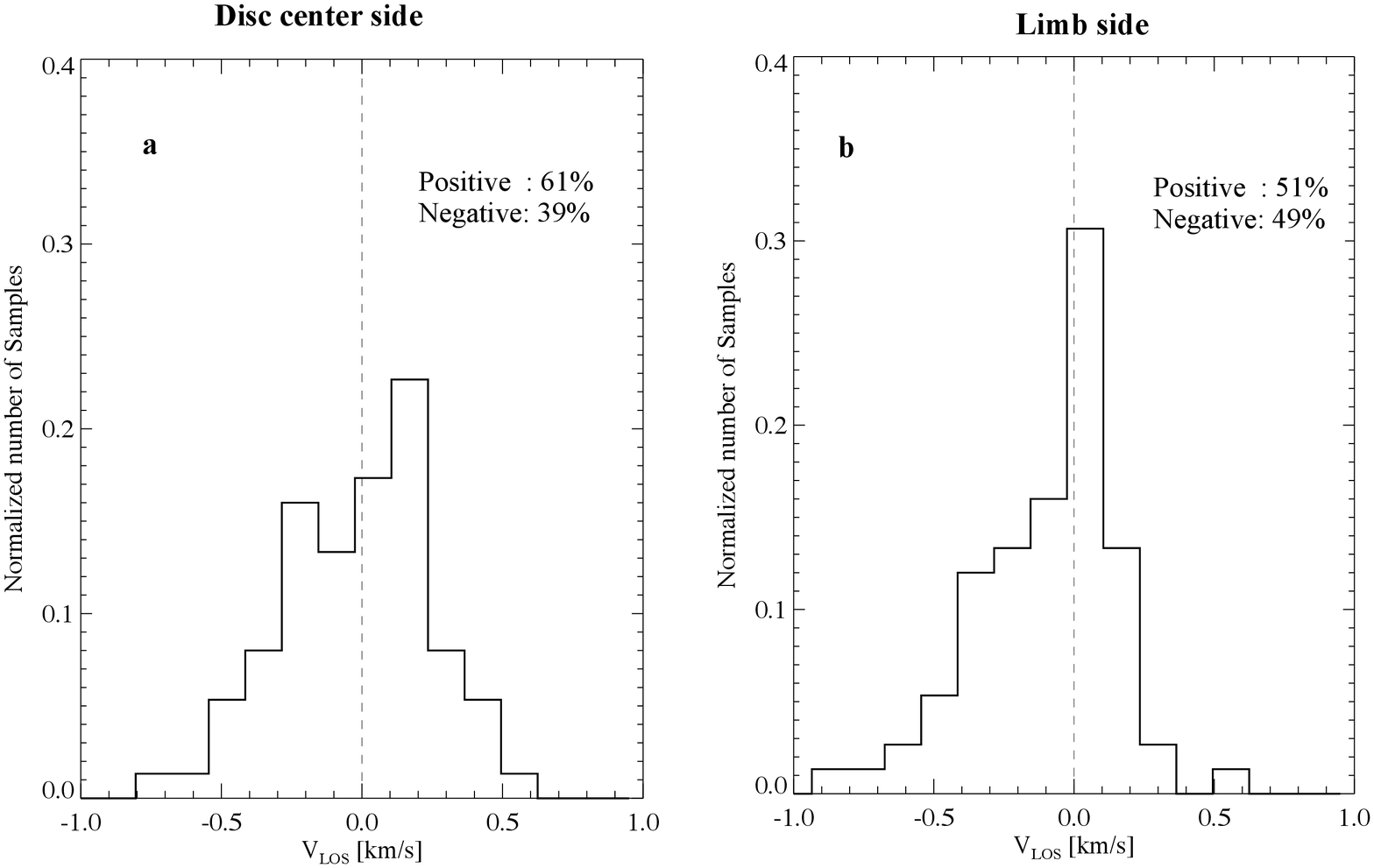}  
    \caption{Same as Figure 15 but for time $t_f$ +16 min.}
\end{center}
\end{figure}

The strength of the converging flow, defined by $V_{\rm r} -  V_{\rm b}$ (see section 3.2.2: Appearance of Magnetic Patches), obtained from the average velocity profiles at $t_0$ - 16 min, $t_0$, $t_f$ and $t_f$ + 16 min is given in Table 2. The converging flow is strong at time $t_0$ and $t_f$ and is weakest at time $t_f$ + 16 minute. This indicates that strong converging flow is necessary to form and maintain magnetic patches
     
\begin{deluxetable}{ccc}
\tabletypesize{\scriptsize}
\tablecolumns{2}
\tablewidth{0pc}
\tablecaption{Variation of Strength of the Converging Flow  with Time}
\tablehead{
\colhead{Time} &\colhead{$V_{\rm r} - V_{\rm b}$}\\
\colhead{[min]} &\colhead{[km/s]}}\\
\startdata

        $t_0$ - 16  & 0.37 \\  

        $t_0$   & 0.52 \\   
   
	$t_f$	  & 0.46 \\  

	$t_f$+ 16	  & 0.2 \\ 

\enddata
\end{deluxetable}

\subsubsection{Case Study}

The evolution of a sample magnetic patch is shown in this section. Top panel of Figure 18 shows the patch evolution in the magnetic flux maps and the bottom panel display the same in normalized continuum intensity maps. The continuum intensity map corresponding to the last detection of the patch (frame 18) shows that a facula is enclosed within the patch and the corresponding magnetic flux map shows that faculae location is cospatial with peak flux location within the patch. The apparent life time of the patch is 32 minutes. The magnetic patch appears to decay via unipolar disappearance. Here we exclude the possibility of flux cancellation with an opposite polarity magnetic patch for the following reasons. Firstly, we do not see an opposite polarity magnetic patch in frame 18 in the vicinity of the patch under consideration, and secondly, we do not think it is probable that an opposite polarity patch appear and cancels the existing patch and both disappear completely within a period of 16 minutes such that no trace is left in frame 19.  
 
 \begin{figure*}[htbp]
 \centering
    \begin{center}
    \includegraphics[width=19cm,height=8cm]{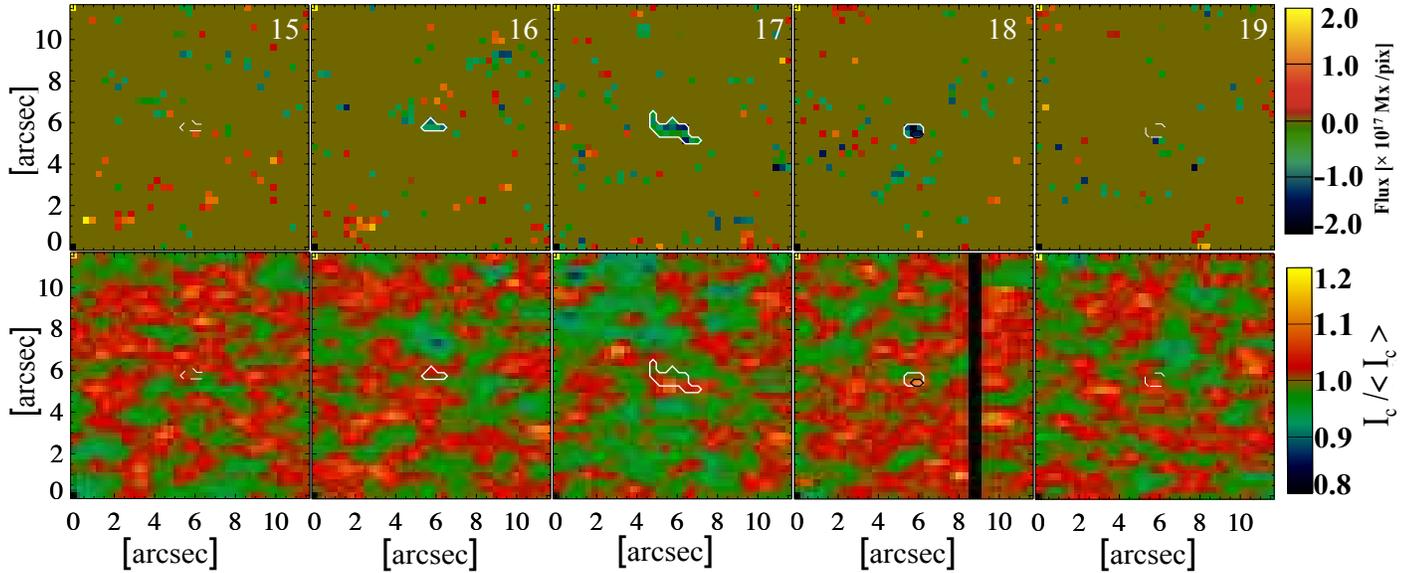}
     \caption{Temporal sequence of the magnetic flux maps (top rows) and normalized continuum intensity maps (bottom rows) showing a sample patch evolution (2013 December 11; South pole). White contours outline the patch, and black contour outline facula. The time interval between consecutive frames is 16 min, and the frame numbers are specified on the top right corner of the magnetic flux maps.}
     \end{center} 
    \end{figure*}

Figure 19 show the variations of magnetic flux and average intensity of the patch with time. The flux evolution displays a rising phase and declining phase. The intensity variation shows that peak average intensity is reached when the patch possesses facula. Throughout its life time the patch has magnetic flux of the order of 10$^{18}$ Mx. However, we detected facula in only one frame. 

\begin{figure}[htbp]
  \centering
  \includegraphics[width=8cm,height=9cm]{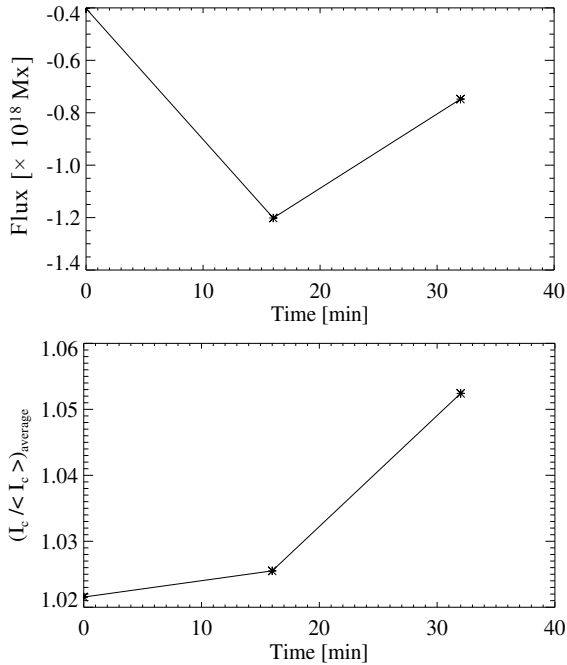} 
     \caption{Top: variation of patch flux with time. Bottom: variation of average intensity of the patch with time.}  
\end{figure}

The photospheric flow in around the patch during its evolution is shown in Figure 20.  The profiles from frame 15 through 18 clearly show the existence of converging flow outside of the patch along the slit direction. The velocity gradient along the slit direction that support convergence is not clear from frame 19. These flow profiles indicate that flow field around individual samples might exhibit a range of variations both in spatial scale and flow strength. 

\begin{figure*}[htbp]
\centering
  \begin{center}
  \includegraphics[width=13cm,height=12cm]{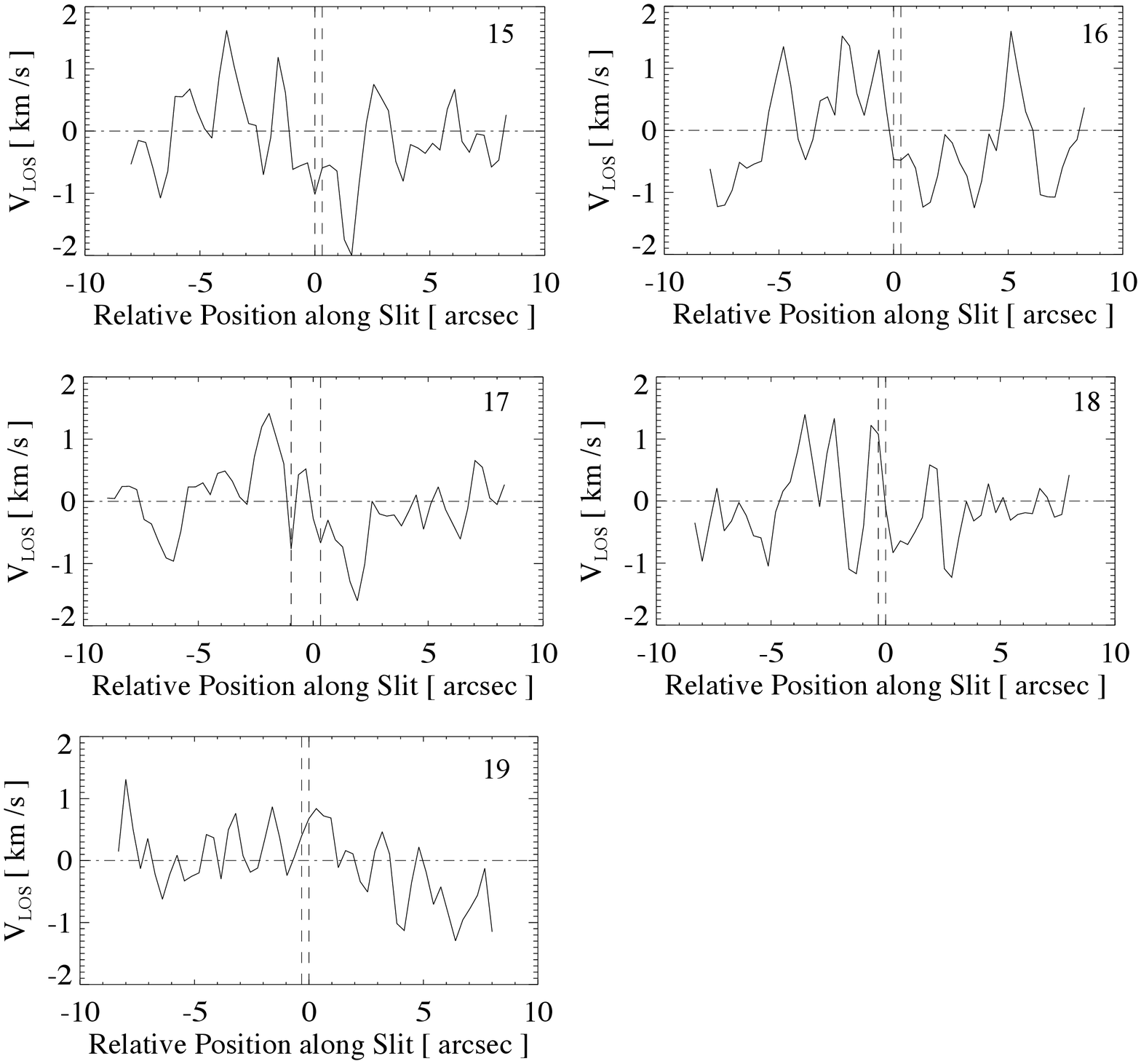} 
     \caption{LOS Doppler velocity profiles displaying flow field in and around the magnetic patch from time$t_0$ - 16 min to $t_f$ +16 min. Limb is towards right.}
      \end{center}  
\end{figure*}


\section{Summary and Discussion}

Our observations correspond to the solar cycle maximum phase. The average life time of the 75 samples used in this study is about 1h. To obtain a statistical distribution of life time of the polar magnetic patches, long duration observations are necessary. \citet{liu} reported that the life time of polar magnetic elements is longer (16.5 h on average) during solar cycle minimum than during cycle maximum (7.3 h). They also found that the dominant polarity elements have longer life time during solar cycle minimum. 

This study present the first observation on the role of converging flow on the formation of the polar magnetic patches, whose flux is believed to originate from the decayed active regions. The uniqueness of the LOS Doppler velocity measurements in the polar region is that the measured velocity is dominated by the horizontal component of the photospheric flow field. Determination of horizontal flow component directly from the Doppler measurements is not possible for disk center quiet Sun; for observations near disk-center, the horizontal velocities are derived using granule tracking technique called Local Correlation Tracking \citep{novb}. The LOS Doppler velocity at any position $(x,y)$ on the solar disk is given by: $V_{LOS}(x,y)= V_{rad}(x,y) \cos\theta + V_h(x,y) \sin\theta$, where $\theta$ is the heliocentric angle, $V_{rad}$ is the radial and $V_h$ is the horizontal component of Doppler velocity at $(x,y)$. The measured Doppler values do not include the component of horizontal flow transverse to the LOS. 

We chose magnetic patches with life time shorter than 6 hours which satisfy a minimum life time criterion of 32 minutes. The $\mu$ value of magnetic centroids of the patches selected for this study fall in the range of 0.3 - 0.2. We found that the polar magnetic patches are surrounded by strong horizontal converging flow during the period of their formation. The converging flow is best represented by the profile at the bisector level 4 which corresponds to a deep photospheric layer. We found from the average LOS velocity profile at the bisector level 4 that peak of the average converging flow velocity is about 0.2 km/s and that its radius of extent is about 3$''$ (after foreshortening correction it is 10$''$ at $\mu$ = 0.3). The strength of LOS velocity around patches was found to increase as one approaches lower photospheric layers, even though the difference in formation height between line core and wing is small. 

Spatial scale after correcting for projection effect and velocity of the converging flow obtained from the average profile at time $t_0$, bear resemblance to that of supergranulation. Supergranular flows are known to concentrate magnetic flux as shown by observations close to the disk center \citep[e.g.,][]{yi}. The supergranulation exhibits fluctuating cell size, diameters of which vary over a wide range between 20 Mm and 50 Mm with an average horizontal length scale of 32 Mm \citep{si}. There are studies which report supergranule cells of much smaller size, with mean diameters between 10 and 20 Mm \citep{hag,ber,sri,der}. Typical horizontal velocity associated with supergranules is in the range of 300-500 m/s \citep{si,sh,hat}; \citet*{sh} also showed that some supergranules are associated with horizontal velocity as high as 1km/s. We think that the dominant contribution to the converging flow comes from the supergranules. However, in terms of dynamic evolution and shorter life-time of the magnetic patches we obtained, it could be possible that a meso-scale flow also is acting on the magnetic patches that influence the local-scale dynamics of the patches.

Similar observational studies have been conducted on solar pores, which are magnetic structures with kG field strengths. \citet{keil} observed that pores form at the supergranular cell boundaries by the advection and concentration of magnetic flux driven by the surface flows. \citet{sob} reported that the horizontal inflows exist and dominate within a 2$''$ zone around pores. Converging flows around pores are also observed by \citet{wang}  and \cite{sa}. 

The mean magnetic flux of the samples considered in this study is about 10$^{18}$ Mx. If we consider impact radius of the converging flow to be about 10$''$ (after foreshortening correction), pixel size of 0.32$''$ and patch flux of 10$^{18}$ Mx then there will be ($0.3\times10^{15}$ Mx) / unit pixel. This calculation indicates that the converging flow we observed is capable of accumulating magnetic flux to form polar patches. We suggest that the isolated 'unipolar appearance' of patches observed in our study most probably occurs due to the coalescence of undetectable flux driven by the converging flow. In view of the dynamic nature of the magnetic patches, we think that in the final stage just before the formation and subsequent evolution of magnetic patches, larger contribution could be coming from flow field in a zonal region with radius smaller than 10$''$. The above calculation also point to the presence of magnetic flux in the polar region that remain invisible even with high resolution observations from Hinode.
 
We also found that horizontal converging flow exist outside of the patches at the time of their apparent disintegration ($t_f$). This suggests that patch decay is not assisted by diverging flows and that magnetic patches are held in place by the incoming flow field. The patch disintegration may not be an instantaneous process. 

Our study is made under the assumption that the contribution from the horizontal component of radial velocity is minor. However, we think it is possible that the dominance of red-shift within the patch at time $t_f$ (Figure 15) is partly contributed by the downdraft within the patch. Assuming that peak down flow speed ($V_{rad}$) due to convective collapse is 5 km/s \citep{nar} and that $V_h$ is negligible during collapse within a magnetic patch, $V_{LOS}(x,y)$ at $\mu$ equal to 0.3 is $\sim$ 1.5 km/s. From the average velocity profile within the patch at time $t_f$ peak velocity is about 0.55 km/s ($V_{LOS}(x,y)$), which when corrected for foreshortening ($\mu$ = 0.3) gives $\sim$ 1.83  km/s. So even if the downdraft speed is smaller than the peak value mentioned above, its contribution is required to account for the preferential redshift Doppler values observed inside the patch at time $t_f$.  

Visual inspection of the samples has shown that some of the magnetic patches die by fragmentation, merging or cancellation and others disappear in isolation (so called 'unipolar disappearance'). It could be possible that in the case of unipolar disappearance, the magnetic patches underwent fragmentation with the fragments that have fluxes below the detection limit of the instrument, thus rendering them invisible. The physical mechanism causing the 'unipolar disappearance' could be one of the possible methods that leads to the reversal of the polar magnetic field. There must be cancellation between opposite polarity magnetic flux fragments happening eventually, at scales invisible to the spatial resolution of Hinode. It would be interesting to obtain the frequency of occurrence of mechanisms, other than direct cancellation with opposite polarity magnetic flux transported from the active latitudes, by which patches die.
 
Unipolar appearance and disappearance of magnetic features are reported also in observations near disc center. \cite{lamb0} noted that the apparent unipolar appearances are due to coalescence of like-polarity diffused magnetic flux using the Hinode-Narrowband Filter Imager (NFI) data. Using the same NFI data set from the quiet Sun \cite{lamb1} investigated death of magnetic features and found that the dominant process by which they die is through flux dispersion (unipolar disappearance).
 
In summary, the patch evolution seems to include three stages, namely, a) the concentration of like-polarity flux fragments by the converging flow field; b) localized concentration of magnetic flux with in the patch which is cospatial with facula; and c) the disintegration of the magnetic patch into like-polarity fragments (see the cartoon below; Figure 21). After Step 3, the magnetic fragments may either coalesce resulting in patch formation or cancel out with opposite polarity fragments. Depending on the phase of the solar cycle one process might dominate over the other; during solar cycle minimum patch formation could be dominating and around cycle maximum cancellation with incoming opposite polarity fragments could be the dominating process.

\begin{figure*}[htbp]
\begin{center}
      \includegraphics[width=10.5cm,height=4cm]{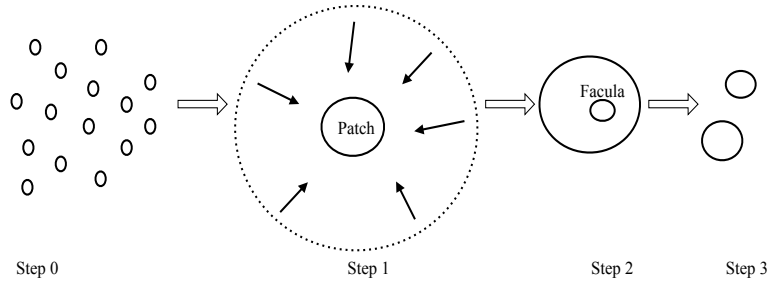}
     \caption{Cartoon showing the evolution of the polar magnetic patch. Step 0: like-polarity flux fragments that are invisible, Step 1: concentration of flux fragments by converging flow into visible large patch, Step 2: faculae appear within the magnetic patch at some moment during the patch evolution, Step 3: disintegration of magnetic patch into flux fragments for recycling or cancellation with opposite polarity fragments in polarity reversal phase.}    
   \end{center}
    \end{figure*}
    
The case study show that the converging flow persists outside of the patch throughout its life time. We also found that not all the patches with magnetic flux of about 10$^{18}$ Mx possess facula. As mentioned in \citet{ajk} observation time could be one of the factors that plays a role in spotting facula inside a patch. Depending on the time of observation we may or may not detect the faculae even if the patch possess enough flux. To understand the physical mechanism behind faculae formation inside the patch and its evolution high spatio-temporal observation and velocity vector information are necessary.

\acknowledgements
Helpful discussions with Drs. D. Orozco Suarez, Robert Cameron, Yukio Katsukawa and Toshifumi Shimizu are gratefully acknowledged. A. J. K. thanks the Ministry of Education, Culture, Sports, Science, and Technology (MEXT) of Japan for financial support through its doctoral fellowship program for foreign students and the GUAS (SOKENDAI) for an associate researcher's grant. \emph{Hinode} is a Japanese mission developed and launched by ISAS/JAXA, with NAOJ as domestic partner and NASA and STFC (UK) as international partners. It is operated by these agencies in cooperation with ESA and NSC (Norway).

\newpage

\bibliography{bib}

\end{document}